\input harvmac
\sequentialequations
%%%%%%%%%%Definitions%%%%%%%%%

\def\np#1{{\it Nucl. Phys.,} {\bf B#1},}
\def\prl#1{{\it Phys. Rev. Lett.,} {\bf #1},}
\def\prd#1{{\it Phys. Rev.,} {\bf D#1},}
\def\plb#1{{\it Phys. Lett.,} {\bf B#1},}

\def\pr#1{{\it Phys. Rev.,} {\bf #1},}
\def\ijmA#1{{\it Int. J. Mod. Phys.,} {\bf A#1},}
\def\rmp#1{{\it Rev. Mod. Phys.,} {\bf #1},}

\def\frac#1#2{\hbox{$#1\over #2$}}
\def\tilde{\widetilde}
\def\hatgama{\widehat\Gamma}
\def\ce{{\cal E}}

\xdef\mysecsym{2.}

\global\newcount\subsecno \global\subsecno=0
\def\subsec#1{\global\advance\subsecno by1
\message{(\mysecsym\the\subsecno. #1)}
\ifnum\lastpenalty>9000\else\bigbreak\fi
\noindent{\it\mysecsym\the\subsecno. #1}\writetoca{\string\quad 
{\mysecsym\the\subsecno.} {#1}}\par\nobreak\medskip\nobreak}

\def\appendix{\global\meqno=1\global\subsecno=0\xdef\secsym{\hbox{A.}}
\bigbreak\bigskip\noindent{\bf Appendix  }
\writetoca{Appendix  }\par\nobreak\medskip\nobreak}

%%%%%%%%%%%%%%references%%%%%%%%%%

\lref\dewit{B.~de Wit, K.~Peeters, and J.~Plefka 1998, ``Superspace Geometry 
for Supermembrane Backgrounds,'' hep-th/9803209.}

\lref\stelle{M.~J.~Duff, and K.~Stelle, 
``Multi-membrane solutions of $D=11$ supergravity,''
\plb{253} 113-118, 1991.}

\lref\duffetal{M.~J.~Duff,  J.~T.~Liu, and J.~Rahmfeld, 
``Dipole Moments of Black Holes and String States,''
\np{494} 161-199, 1996, hep-th/9612015.}

\lref\kallosh{R.~Brooks, R.~Kallosh, and T.~Ortin,
``Fermion Zero Modes and Black Hole Hypermultiplet with Rigid Supersymmetry,''
\prd{52} 5797-5805, 1995, hep-th/9505116.}

\lref\embacher{P.~C.~Aichelburg, and F.~Embacher,
``Exact superpartners of N=2 supergravity solitons,''
\prd{34}, 1986, 3006-3011.}

\lref\schwinger{W.~Rarita, and J.~Schwinger,
``On a Theory of Particles with Half Integral Spin,'' \pr{60} 61-61, 1941.
\hfill\break
See also section 5.6 of  S.~Weinberg 1995, {\it The 
Quantum Theory of Fields}, Volume I (Cambridge: Cambridge University Press).
}

\lref\gueven{R.~G\"uven, ``Black p-brane solutions of D=11
supergravity theory,''
 \plb{276} 49-55, 1992.}

\lref\witten{E.~Witten, ``IS SUPERSYMMETRY REALLY BROKEN?''
\ijmA{10} 1247-1248, 1995, hep-th/9409111.}

\lref\becker{K.~Becker, M.~Becker, and A.~Strominger,
``THREE-DIMENSIONAL SUPERGRAVITY AND THE COSMOLOGICAL CONSTANT,''
 \prd{51} 6603-6607, 1995, hep-th/9502107.}

\lref\deser{S.~Deser, J.~H.~Kay, and K.~S.~Stelle, ``Hamiltonian
Formulation of Supergravity,'' \prd{16} 2448, 1977.}

\lref\weinberg{S.~Weinberg, ``The Cosmological Constant Problem,''
\rmp{61} 1-23, 1988.}

\lref\win{V.~Balasubramanian, D.~Kastor, J.~Traschen, and K.~Z.~Win,
 hep-th/9811037.}

\lref\bandos{I.~Bandos, N.~Berkovits, and D.~Sorokin,
``Duality-Symmetric Eleven-Dimensional Supergravity and its Coupling
to M-Branes,'' \np{522} 214-233, 1998, hep-th/9711055.}

\lref\zumino{B.~Zumino, ``SUPERSYMMETRY AND THE VACUUM,'' \np{89} 
535-546, 1975.}

\lref\townsend{G.~W.~Gibbons, and P.~K.~Townsend, 
``Vacuum Interpolation in Supergravity via Super $p$-branes,'' \prl{71}
3754-3757, 1993.}

\lref\gibbons{G.~W.~Gibbons, in {\it Supersymmetry, Supergravity and
Related Topics}, edited by F.~Del~Aguila, J.~A.~de.~Azc\'arraga, and 
L.~E.~Iba\~nez (World Scientific, Singapore, 1985).}

\lref\sorokin{D.~Sorokin, and P.~Townsend, ``M-theory superalgebra from the 
M-5-brane,'' \plb{412} 265-273, 1997, hep-th/9708003.}

\lref\claudio{S.~Deser, and C.~Teitelboim, ``Supergravity Has
Positive Energy,'' \prl{39} 249-252, 1977.}

%%%%%%The End of References%%%%%%%%%

%%%%%%%%%% Title Page %%%%%%%

\Title{\vbox{\baselineskip12pt
\hbox{hep-th/9811082}
}}
{\vbox{\centerline{\titlerm Fermion Zero Modes and Cosmological Constant}
 }}

\centerline{ K.~Z.~Win  }
\bigskip
\centerline{\it Department of Physics and Astronomy}
\centerline{\it University of Massachusetts, Amherst, MA 01003, USA}
\bigskip

\bigskip
\centerline{\bf Abstract}
\bigskip
A general condition for the existence of  fermion zero modes is 
derived for the M-5-brane, the M-2-brane and the 
$D=4$, $N=2$ Majumdar-Papapetrou
0-brane. The fermion zero modes of these $p$-branes do not exist 
if the
supersymmetry spinor generator goes to a constant at the
horizon and they
exist only if it vanishes there.  In particular 
it is  shown that the  fermion zero mode of the M-2-brane in $D=11$ 
can be forbidden from existence if Rarita-Schwinger gamma tracelessness
condition is imposed  on the gravitino field.  Non-existence of fermion 
zero mode  is interpreted,
in analogy to the three dimensional example of Becker et.~al., 
 as  a world with  zero cosmological constant  
without supersymmetric excited states. 
Also derived are the spin of the M-5-brane
 and its 3-form electric and magnetic dipole
 moments.

%\draft\vfill\eject
\Date{November 1998}
%%%%%%%%%%%%%%%

\newsec{Introduction}  
\noindent
The cosmological constant problem in a supersymmetric theory is to 
reconcile the observed zero cosmological constant  with the observed
broken supersymmetry of the world. Witten \witten\ has made a general
observation that in three 
dimension the 
 cosmological constant can be zero without the  
supersymmetric multiplet of physical  states
due to conically singular geometry
 of any massive three dimensional spacetimes.
Three dimensional $N=2$ supersymmetric 
abelian Higgs model coupled to supergravity has since 
been studied by Becker, Becker, and Strominger \becker\ 
as an  evidence of Witten's claim.  Their calculation involves 
proving that the vortex soliton solution of the model preserves 
some  supersymmetries and yet fermion zero mode does not exist since
the gravitino is not normalizable.

This paper  will show that analogous results to that of 
Becker, Becker, and Strominger can be obtained in higher 
dimensional supergravities.  In particular we will study
the existence of the fermion zero modes of  2-brane and 
5-brane solitons in $D=11$ and show that for some functional choice  of the  
 spinor generator fermion zero modes can be forbidden from existence. 

The first study
of the supergravity fermion
 zero modes  was done by Aichelburg and Embacher \embacher\ 
 for the
$D=4$, $N=2$ Majumdar-Papapetrou 0-branes
and the existence of the fermion zero modes was later  related \kallosh\
to the
Rarita-Schwinger (RS) condition $\Gamma^m\psi_m=0$.
We will re-examine the existence of the
fermion zero modes in this theory, and 
show that  the RS condition is not a necessary condition but merely a
sufficient 
($D=4$) condition  for the
existence of the fermion zero modes.

More importantly we show that the fermion zero modes of
$D=11$ M-2-brane \win\  in fact do not  exist  if the RS condition is 
imposed i.e.~it
 is not even a sufficient condition (for the M-2-brane.)
This opens up the possibility that the RS condition needs not be imposed 
on the fermion zero modes in general.   In fact it is known \embacher\
that the RS condition is not preserved at the second order
in spinor parameter if imposed at the first order. 
In their original work \schwinger\ 
the  condition of Rarita and Schwinger is only one of a number 
of conditions imposed on $\psi_m$ in order to make it behave as
a spin $\frac 32$ four dimensional particle.  But in the context 
of constructing fermion zero modes described in this paper there
 seems to be absolutely no reason to impose this condition.
 We will
argue that the RS condition should not be  related to the existence of the
fermion zero modes.  Instead the desired existence will be shown to
be  related to the value of the supersymmetry generator at the horizon.
 The RS 
condition does fix that value but it is not the unique way.

The fermion zero modes exist if the gravitino is normalizable.
The normalizability of the gravitino will be shown to be related to the 
behavior of $\epsilon$ at $r\to 0$ i.e.~at the horizon.\foot{We
 will use the isotropic coordinates so that $r\to 0$ is the horizon limit. 
$r$ is the radial distance transverse to the world volume.}
All fermion zero modes (if they exist) are equivalent 
if $\epsilon$ goes to the same constant as $r\to\infty$ but, as we will see,
the gravitino is normalizable if and only if 
$\epsilon$ vanishes as $r\to 0$.  In other words the supersymmetry
must be completely broken at the horizon for the existence of the fermion 
zero mode!
So long as 
$\epsilon\sim r^{-\delta}$ ($\delta\ge 0$)
as $r\to 0$, the normalizability of $\psi_m$ is forbidden.
This general result will be derived for the $D=4$, $p=0$ case and
 the M-2-brane  as well as for the M-5-brane case.
For the M-2-brane case the RS condition  makes $\epsilon=$constant and as
a result the normalizability of $\psi_m$ is forbidden 
in this case. 

It has been shown by Gibbons \gibbons\ for the Majumdar-Papapetrou 
 0-brane 
and more recently by Gibbons and
Townsend \townsend\ for the M-$p$-branes that all $p$-branes of interest here
interpolate between two maximally supersymmetric vacua: ${\cal M}_D$
 at $r\to\infty$
and $AdS_{p+2}\times S^{D-p-2}$ at
$r\to 0$. The vacuum at infinity has zero cosmological constant\foot{This 
is a classical result but quantum Casimir energies cancel between equal
contributions of  boson and fermion loops
as explained
in \weinberg\ and first derived in \zumino.  Such arguments are still
valid because these $p$-branes  preserve some supersymmetry.}
 and only if the supersymmetry
is completely broken at the horizon fermion zero modes exist!  The 
generalization of the observation by Gibbons and
Townsend  is then, for $\epsilon{\buildrel r
\to 0\over\longrightarrow}0$ case, 
these $p$-branes interpolate between ${\cal M}_D$
with supersymmetric excited states and $AdS_{p+2}\times S^{D-p-2}$
 with zero supersymmetry generator.   For the 
 alternative behavior of $\epsilon{\buildrel r\to 0\over
\longrightarrow}$constant the $p$-branes 
interpolate between ${\cal M}_D$ 
without fermion zero mode and $AdS_{p+2}\times S^{D-p-2}$
 with nonzero supersymmetry generator.
There is then a sort of 
duality between  zero cosmological constant universe
with nonsupersymmetric states and a negative cosmological constant world
with nonzero supersymmetric generator and  vice versa.

It should be emphasized that the present paper  does not claim to 
solve the cosmological constant problem \weinberg\ in all its gory
details. The models presented here,
consistent as they are, 
may or may not correspond to the reality.  It also raises 
a question of why  the amount of supersymmetry at the horizon is unbroken in 
such a 
way that the observed nonsupersymmetric universe has zero cosmological 
constant.
But the question is
no different from question such as why the observed four dimensional
universe is compactified, in some very
 special way, from a higher dimensional manifold. Anthropic considerations 
may or may not have the answer.

The  main results of this paper are the connection between
the supersymmetry breaking at the horizon and the existence (or 
nonexistence) of the fermion zero modes and  extending the results of 
reference \becker\ to higher dimensional theories.
In  section 2 we illustrate the generation of fermion zero modes 
using the M-5-brane as an example. We will see that the result is a spinning 
M-5-brane, called fivebrane superpartner,
 with two dipole moments of 3-form gauge field.  
We then derive the conditions for the existence of the fermion zero modes
for three $p$-branes.

\newsec{Fivebrane Superpartner}
\noindent
We adopt 
the eleven dimensional convention of \dewit\ and  split
 eleven coordinates $(x^m=x^0,x^+,x^-,x^8,x^7,x^6,x^1,\dots,
x^5)$
 into $p+1$ coordinates $x^a$
tangent to the world volume of the M-$p$-brane and $D-p-1$
 coordinates $x^\alpha$
transverse to the world volume.
For the case of $p=5$ we will take $(x^a=x^0,x^+,x^-,x^8,x^7,x^6)$ and 
$(x^\alpha=x^1,\dots,x^5)$.  For the $p=2$ case to be discussed later
it will be understood that there are  three $x^a$ and eight $x^\alpha$ but 
it will not be necessary to say which of the spatial $x^m$ belong to $x^a$.
The bosonic 5-brane solution of G\"uven \gueven, described by the
following
fields 
\eqn\fivebrane{
ds^2={\eta_{ab}\over 
f}dx^adx^b+f^2\delta_{\alpha\beta}dx^\alpha
dx^\beta \>,\quad F_{\alpha_1\dots\alpha_4}
=-s\varepsilon_{\alpha_1\dots\alpha_4}{}^{\alpha}\partial_{\alpha}
(f^3)\>,\quad\psi_m=0
}
is invariant under the following 
eleven dimensional supersymmetry transformation
\eqnn\susyA
\eqnn\susyB
$$\eqalignno{
\delta\psi_m=&\left[\partial_m+
\frac 14 \omega_m^{\hat n\hat p}\hatgama_{np}+
\frac 1 {288} \left(\Gamma_m{}^{npqr}-8\delta_m^n\Gamma^{pqr}\right)
F_{npqr}\right]
\epsilon &\susyB\cr
 \delta A_{mnp}=&-6\bar\epsilon\, \Gamma_{[mn}\psi_{p]},\hskip4mm
\delta e_m^{\hat n}=2\bar\epsilon\,\hatgama^n\psi_m &\susyA\cr}
$$
where $\epsilon$ is an
 anticommuting Majorana spinor, $s=\mp 1$ the sign of
the 
fivebrane magnetic charge, $\varepsilon_{12345}=\varepsilon^{12345}=1$,
$\{\hatgama^m,\hatgama^n\}=2\eta^{mn}$, and $\Gamma_m=e_m^{\hat n}\hatgama_n$.
  The hatted quantities are
flat space ones and $\Gamma$ with multiple indices  product
of dirac matrices with all indices different.
The  supertorsionless equation consistent with equation \susyB\
is $de^{\hat m}=e^{\hat n}\wedge \omega^{\hat m}{}_{\hat n}+$ terms
involving $\psi$ (which we do not need here.) 
The function $f=f(x^\alpha)$ satisfies $\delta^{\alpha\beta}\partial_\alpha
\partial_\beta f^3=0$ which solution of interest is
\eqn\harmonic
{
f=\left(1+{6M\over |\vec r|^3}\right)^{\frac 13}
}
where $\vec r=(x^1,\dots,x^5)$. 
 Invariance of the solution  \fivebrane\
under \susyA\ is easy to see. But the
 solution  \fivebrane\ is invariant \gueven\ 
under \susyB\
only if $\epsilon=f^{-\frac 14}\lambda$ with constant $\lambda$ satisfying
$(1+\tilde\Gamma s)\lambda=0$ where $\tilde\Gamma\equiv\hatgama^{12345}$.
Half of the independent
 components of $\lambda$ are zero because $\tilde\Gamma$ squares to unity
and $\tr\tilde\Gamma=0$.  The fivebrane solution therefore preserves half
of the supersymmetry.  

The fermion zero mode of the fivebrane \fivebrane\ is obtained by acting on 
the solution with a spinor $\epsilon=\ce\lambda$
 with the property 
\eqn\projection
{
(1-\tilde\Gamma s)\lambda
=0
}
 where we have placed all  spacetime dependence of $\epsilon$ in 
the function $\ce=\ce(x^m)$.  From equation \susyB\ we will then generate
nonzero $\psi_m$ which in turn can be plugged into equation \susyA\ 
to generate the fermionic corrections to the bosonic fields.  Equations of
motions of the supergravity are invariant under equations \susyB\ and \susyA\
with $\epsilon$ any  function of spacetime coordinates but 
the resultant fermion zero mode will have the desirable physical 
properties only if we take $\ce$ to be a function of the transverse
spatial coordinates with the property $\ce\> {\buildrel r\to\infty\over 
\longrightarrow}$ a nonzero constant.  This property is necessary in order to obtain the asymptotically flat spacetime and amounts to restricting local 
supersymmetry.  By rescaling $\lambda$ we may set the asymptotic value
of $\ce$ to 1.  These properties of $\ce$ are obtained by taking
\foot{We discuss a more general form of  ${\ce}$ in section 4.}
$\ce=f^{-\delta_5}$.
  In this section we will assume that the value
of $\delta_5$ is such that the resultant gravitino is normalizable. For 
the most part in this section we
will not need to know the precise value of $\delta_5$ since we will mainly 
be
interested in $r\to\infty$ behavior of various quantities.   Also,  as
explained in \win, the interactions between two 
superpartners are well defined only at distance far away from the horizon.

We define $\Lambda_{mn\dots}\equiv\bar\lambda\hatgama_{mn\dots}\lambda$
and raise and lower indices of $\Lambda$ with the Minkowski metric.
Carrying out the procedure described in the previous paragraph we get the
$\lambda^2$ quantities to be
\eqnn\gravitino
\eqnn\vielbein
\eqnn\electric
\eqnn\magnetic
\eqnn\metric
$$\eqalignno{
\psi=&\left\{\half f^{-\frac 52}f_\alpha\hatgama^\alpha{}_adx^a-{1\over f}
\left[f_\beta\hatgama^\beta{}_\alpha+f_\alpha(\delta_5+\frac 14)\right]
\right\}\epsilon &\gravitino\cr
A_{a_1a_2 a_3}\sim & {3Qx^\beta\varepsilon_{a_1a_2a_3}
{}^{b_1b_2 b_3}\Lambda^{\beta}{}_{b_1b_2 b_3}\over 2r^5}&\electric\cr
A_{c\alpha_1\alpha_2}\sim &
{9 Qx^{\gamma}{}_{\alpha_1\alpha_2\gamma}{}^{\beta_1\beta_2}
\Lambda_{c\beta_1\beta_2}\over 2r^5}&\magnetic\cr
e^{\hat a}\sim & {6 Mx^\beta\Lambda^a{}_{\beta\alpha}\over r^5}dx^\alpha\>,
\qquad  e^{\hat \alpha}\sim {3 Mx^\beta\Lambda^\alpha{}_{\beta a}\over
r^5}
dx^a &\vielbein\cr
g_{\alpha c}\sim &{9Mx^\beta\Lambda_{\beta\alpha c}\over r^5}&\metric
}
$$
where $f_\alpha=\partial_\alpha f$, $Q\equiv sM$
 the magnetic charge, and $\sim$ means
$r\to\infty$ limit.
We have also used the conventions 
 $\varepsilon_{0+-876}=\varepsilon^{0+-867}=1$,
 and $\hatgama^{0+-87612345}=1$.
 Other $\lambda^2$  components of the bosonic fields 
are zero on account of the Majorana property of $\lambda$.  

The ``generalized'' spin $J$ of the M-5-brane can be read off from the off-diagonal components of
the metric \metric\ as follows
\eqn\spin
{
g_{\alpha c}\sim {x^\beta J_{\beta\alpha c}\over r^5}\> \Longrightarrow
J_c{}^{\alpha\beta}=9M\Lambda_c{}^{\alpha\beta}.
}
We will call the components $A_{a_1a_2 a_3}$ electric and the electric
dipole moments $P$ can be read off  by defining
\eqn\edipole
{
A_{a_1a_2 a_3}\sim {x^\beta\varepsilon_{a_1a_2 a_3}{}^{b_1b_2 b_3}
P_{\beta b_1b_2 b_3}\over 6r^5}\> \Longrightarrow P^{\beta}{}_{abc}
=9Q\Lambda^\beta{}_{abc}
}
As in 3+1 dimensional electrodynamics,
the electric dipole moments transform as  vectors in the
 transverse space if the
tangent space indices are considered mere labels of the vectors.  There
are thus 20 electric dipole moments.
We will call the remaining components of the gauge field magnetic and 
the magnetic dipole moments $\mu$ can be read off by defining
\eqn\mdipole
{
A_{c\alpha_1\alpha_2}\sim {x^{\gamma}\varepsilon_{\alpha_1\alpha_2\gamma}
{}^{\beta_1\beta_2}\mu_{c\beta_1\beta_2}\over 2r^5}\>
\Longrightarrow \mu_c{}^{\alpha\beta}=9Q\Lambda_c{}^{\alpha\beta}
}
As in 3+1 dimensional electrodynamics,
the magnetic dipole moments transform as antisymmetric tensors
in the transverse space.\foot{Of course in the elementary exposition of 
3+1  electrodynamics  magnetic dipole is a (pseudo-)vector which in 3+1 
dimension
is an antisymmetric tensor in disguise.}
 Six tangent space indices can be considered
as labels of these tensors.  

Because $\mu$ and $J$ have the same index structure it is possible to 
calculate the gyromagnetic ratio from the naive formula 
$$
\mu={\rm g}{Q\over 2M}J
$$
to get g=2.  But this value will depend sensitively on the definition 
\mdipole. It should be compared with the usual g in $D=4$, $p=0$ case
only after careful comparison of two theories which we have not done here.

It is  surprising that a spinning purely magnetic object can have dipole
moments other than the electric ones.  Some physicists \duffetal\
have already observed an analogous phenomena in four dimensions 
where they interpret a similar result 
 as spinning electric charge with both
electric and magnetic dipole moments.  We remark that 
the recent result \sorokin\
of M-5-brane superalgebra including both 2-form and 5-form 
charges has been interpreted in \bandos\ as M-5-brane being dyonic.
Perhaps  these objects are  composite like
the garden variety neutron which has zero {\it net} electric charge as well
as highly
nontrivial magnetic dipole moment.

\newsec{The Normalizability of $\psi_m$}
\noindent
The norm of the gravitino is by definition 
\eqn\norm
{
\|\psi\|^2=\int{\psi_m}^\dagger\psi_ng^{mn}
}
where the integral is over all of the transverse space from horizon
to infinity. 
We will always work in spherical coordinates.
Above $g^{mn}$ is the background metric.

\subsec{M-5-brane}
\noindent
From equation \gravitino\ we get
$$\eqalign{
\|\psi\|^2=&\Omega_4\int_0^\infty r^4f^5{\psi_m}^\dagger
\psi_ng^{mn}\, dr=36
[5+(\delta_5+\frac 14)^2]M^2\Omega_4\lambda^\dagger\lambda
\int_0^\infty f^{-(2\delta_5+3)}{dr\over r^4}\cr
=&12[5+(\delta_5+\frac 14)^2]M\Omega_4\lambda^\dagger\lambda
\int_0^\infty(1+6t)^{-(\frac {2\delta_5}3+1)}dt\cr
=&12
[5+(\delta_5+\frac 14)^2]\Omega_4\lambda^\dagger\lambda
\cases{-(1/4\delta_5)(1+6\infty)^{-{2\delta_5\over 3}},
&if $\delta_5<0$;\cr
(1/6)\ln(1+6\infty),&if $\delta_5=0$;\cr
(M/ 4\delta_5), &if 
$\delta_5>0$}
}$$
where $\Omega_z=2\pi^{z/2}/(\frac z2-1)!$. 

Imposing the RS condition we have
\eqnn\rarita
$$\eqalignno{
\Gamma^m\psi_m=&\left[
 \half f^{-2}f_\alpha\hatgama^a\hatgama^{\alpha}{}_a
+f^{- 2}f_\beta\hatgama^\alpha\hatgama_\alpha{}^\beta -
f^{- 2}(\delta_5+\frac 14)\hatgama^\alpha f_\alpha\right]\epsilon\cr
=& f^{- 2}f_\alpha\hatgama^\alpha(-3+4-\delta_5-\frac 14)
\epsilon= 0&\rarita\cr
\Longrightarrow &\delta_5=\frac 34
}
$$
where we have used an obvious identity $\hatgama_\alpha{}^\beta
=\hatgama_\alpha
\hatgama^\beta-\delta_\alpha^\beta$.
Therefore the RS condition is consistent with the
finiteness (and positivity) of  the gravitino norm.

\subsec{M-2-brane}
\noindent
Recall that in this subsection  the 
$a$ type indices take on three values whereas the $\alpha$ type
indices take on eight spatial values. 
 The gravitino field derived in \win\ can be easily 
generalized for $\epsilon
=f^{-\delta_2}\lambda$ with constant $\lambda$ satisfying an analogous 
condition to equation \projection.  The result is
\eqn\psitwo
{
\psi =\left\{
f^{-\frac 52}f_\alpha\hatgama^{\alpha}{}_adx^a
-{1\over 2f}\left[\hatgama^\beta{}_\alpha f_\beta+(2\delta_2+1)f_\alpha
\right]dx^{\alpha}\right\}\epsilon
}
where the background metric and the function $f=f(x^\alpha)$ are given by
\stelle\
$$
ds^2={\eta_{ab}\over
f^2}dx^adx^b+f\delta_{\alpha\beta}dx^\alpha dx^\beta,\qquad
f=\left(1+{3M\over r^6}\right)^{\frac 13},\qquad r=\sqrt{\delta_{\alpha
\beta}x^\alpha x^\beta}\>.
$$
We have
$$
\eqalign{
\|\psi\|^2=&\Omega_7\int_0^\infty r^7f^4{\psi_m}^\dagger\psi_ng^{mn}\, dr=
9\lambda^\dagger\lambda M^2[11+(2\delta_2+1)^2]\Omega_7\int_0^\infty
f^{-(2\delta_2+3)}{dr\over r^7}\cr
=&\frac 32\lambda^\dagger\lambda M \Omega_7[11+(2\delta_2+1)^2]
\int_0^\infty(1+3t)^{-(\frac {2\delta_2}3+1)}\, dt\cr
=&\frac 32 \lambda^\dagger\lambda \Omega_7[11+(2\delta_2+1)^2]
\cases{-(1/2\delta_2)(1+3\infty)^{-{2\delta_2\over 3}},
&if $\delta_2<0$;\cr
(1/3)\ln(1+3\infty),&if $\delta_2=0$;\cr
(M/ 2\delta_2), &if 
$\delta_2>0$.}
}
$$
Imposing the RS condition we have
$$
\eqalign{
\Gamma^m\psi_m=&\left[f^{-\frac 32}f_\alpha\hatgama^a\hatgama^\alpha{}_a
-f^{-\frac 32}f_\beta\hatgama^\alpha\hatgama^\beta{}_\alpha
-f^{-\frac 32}(2\delta_2+1)\hatgama^\alpha f_\alpha\right]\epsilon\cr
=&f^{-\frac 32}f_\alpha\hatgama^\alpha(-3+\frac 72-\delta_2-\half)\epsilon=0\cr
\Longrightarrow &\> \delta_2=0.
}
$$
We see that RS condition with $\delta_2=0$ forbids the existence of the
fermion zero mode.

\subsec{Majumdar-Papapetrou 0-brane}
\noindent
We will follow the metric signature and the supersymmetry transformation
equations  
 of \embacher\ but three
indices $\alpha,\beta$ will  
denote the usual three dimensional spatial indices and $m,n$
four spacetime coordinates.  The background
metric and the harmonic function are
$$
ds^2={1\over f^2}dt^2-f^2d\vec x^2\>\qquad f=
1+{M\over r}
$$
Taking $\epsilon=f^{-\delta_0}\lambda$ with $\lambda$ satisfying an
analogous relation to \projection\ we can straightforwardly generalize
the gravitino given in \embacher\ (or \kallosh) to get
\eqn\psifour
{
\psi=\left\{
-{f_\beta\over f^3}\hatgama^\beta dt
-{1\over f}\left[\hatgama^\beta\hatgama_\alpha 
f_\beta+(\delta_0-\half)f_\alpha\right]
dx^\alpha\right\}\epsilon.
}
As shown in \embacher\ the RS condition is equivalent to $\delta_0=\half$.
Bearing in mind that, in this subsection, 
 the spatial dirac matrices are negative
complex transpose of themselves
we 
can derive
$$
\psi^\dagger=\epsilon^\dagger\left\{
{f_\alpha\over f^3}\hatgama^\alpha dt
-{1\over f}
\left[\hatgama_\alpha\hatgama^\beta
f_\beta+(\delta_0-\half)f_\alpha\right]
dx^\alpha\right\}
$$
We next calculate
${\psi_n}^\dagger\psi^n
$ and  we get
$$
\eqalign{
{\psi_m}^\dagger\psi^m=&{1\over f^4}\epsilon^\dagger \Big[
-f_\beta f_\alpha\hatgama^\alpha\hatgama^\beta
-(\delta_0-\half)^2
\delta^{\alpha\beta}f_\alpha f_\beta\cr
&+f_\beta f_\gamma\hatgama_\alpha
\hatgama^\beta\hatgama^\gamma\hatgama^\alpha
+
2(\delta_0-\half)\hatgama^\beta\hatgama^\alpha f_\alpha f_\beta\Big]\epsilon
\cr
=&{1\over f^4}\epsilon^\dagger \bigg\{
{M^2\over r^4}\left[1-(\delta_0-\half)^2-2(\delta_0-\half)\right]
-f_\gamma f_\beta\hatgama_\alpha\hatgama^\beta\left(\hatgama^\alpha
\hatgama^\gamma+2\delta^{\alpha\gamma}\right)\bigg\}\epsilon\cr
=&{1\over f^4}\epsilon^\dagger \bigg\{
{M^2\over r^4}\left[2-(\delta_0+\half)^2\right]+f_\gamma f_\beta
\left(\hatgama^\beta\hatgama^\gamma-2\hatgama^\gamma\hatgama^\beta\right)
\bigg\}\epsilon\cr
=&f^{-(4+2\delta_0)}\lambda^\dagger\lambda
\left[3-(\delta_0+\half)^2\right]M^2/r^4
}
$$
We finally get
$$
\eqalign{
\|\psi\|^2=& \Omega_3[3-(\delta_0+\half)^2]\lambda^\dagger\lambda
 M^2\int_0^\infty f^{-(1+2\delta_0)}\, {dr\over r^2}\cr
=&\Omega_3[3-(\delta_0+\half)^2]\lambda^\dagger \lambda
M\int_0^\infty (1+t)^{-(1+2\delta_0)} \, dt\cr
=& \Omega_3[3-(\delta_0+\half)^2]
\lambda^\dagger \lambda\cases{-(1/2\delta_0)(1+\infty)^{-2\delta_0}
,&if $\delta_0<0$;\cr
\ln(1+\infty), &if $\delta_0=0$;\cr
(M/2\delta_0), &if $\delta_0>0$.}
}
$$
We see that the RS condition with $\delta_0=\half$ is only a sufficient
condition for the normalizability to the norm.  In contrast to the
eleven dimensional cases the positivity of the four dimensional 
norm sets an upper bound on $\delta_0$: $\delta_0<\sqrt{3}-\half$.
For the ease of exposition we will have this bound in our mind whenever
we speak of $\delta_0$ without mentioning the bound.

\newsec{Summary}
\noindent
We have shown that the normalizability of the 
gravitino can be achieved if $\delta_p>0$ for all three $p$'s.   Because
the dummy
variable of integration is $t=r^{3-p-D}$ the divergence 
of the  
integral is due to   the behavior of  
$\epsilon$ at $r=0$.  From the fact that   $\delta_p=0$ gives 
rise to logarithmic divergences, 
it
is easy to convince oneself that   $\delta_p$ dependence of the 
normalizability 
  will be unchanged if
one has only demanded that 
$\epsilon{\buildrel r\to 0\over\longrightarrow}r^{\delta_p}$ regardless
of the behavior of  $\epsilon$ elsewhere.\foot{Of course
$\epsilon$ must not behave wildly elsewhere.} Put it another way {\it the
 norm exists if and only if $\epsilon$ vanishes at the horizon.}  We can 
take 
the vanishing of $\epsilon$ at the horizon as the complete breaking
of supersymmetry there.
On the other hand  {\it the norm diverges logarithmically
 if $\epsilon$ goes to 
a constant at the horizon} and we can take constant $\epsilon$
at the horizon as some preserved 
supersymmetry there.  Note that 
the logarithmic behavior of the divergence due to the infrared limit
is the same as that observed
in three dimensional example \becker.

The existence of the norm is related to the existence of the quantum
fermion states.  Although there is no well defined  notion of the
classical 
fermion  one can still ask in what sense the gravitino
is well behaved classically. Because they depend on the
coordinate system  the 
values of $\psi_m$ have no meaning.   The correct things to examine are
$\psi_{\hat m}$.
In fact in the Hamiltonian formulation \deser\ $\psi_{\hat m}$ are the
fundamental quantities. It is trivial to check that
$\psi_{\hat m}$ in all three cases are nonsingular at the
horizon so long as  $\delta_p\ge 0$.
Therefore  the gravitino is well behaved in the  
classical sense for constant  $\epsilon$ at the horizon
  despite the nonexistence of the quantum norm in this case.

For completeness we finally
 note that Deser and Teitelboim
 \claudio\ had proposed the so called ``natural'' condition 
$\Gamma^\alpha\psi_\alpha=0$ in the context of defining
the supercharge.  This ``natural'' condition turns out to be equivalent to 
$\delta_0=\frac 32, \delta_2=3, \delta_5=3\frac 34$.

\medskip
I acknowledge discussions with David Kastor.

\listrefs
\bye